\documentclass[pre]{revtex4}
\usepackage{graphics}
\begin{document}
\title{Path-integral approach to the dynamics in sparse
random network}
\author{Takashi Ichinomiya}
\affiliation{Laboratory of Nonlinear Studies and Computation, Research Institute for
Electronic Science, Hokkaido University, Sapporo, Hokkaido, Japan}
\email{miya@aurora.es.hokudai.ac.jp}
\begin{abstract}
 We study the dynamics  involved in a sparse random network model. We
 extend the standard  mean-field approximation for  the dynamics of a
 random network by employing the path-integral approach.
 The result indicates that the distribution of the variable is essentially
 identical to that  obtained from globally coupled oscillators with
 random Gaussian interaction. We present the results of a numerical
 simulation of the  Kuramoto transition in a random network, which is
 found to be consistent  with this analysis.
\end{abstract}
\date{\today}
\maketitle
\section{Introduction\label{introduction}}

 Many systems in nature, such as food webs, metabolic systems,
 coauthorship of papers, the worldwide web, and so on, can be
 represented as complex networks\cite{Watts, Barabashi-Review, Newman,
 Dorogovstev}. Investigations of real networks have shown  that these networks
 have topologies  different from random networks. In particular, we have
 recognized that many networks have scale-free degree distribution,
 $P(k)\propto k^{-\gamma}$, where $k$ is the degree of nodes.

 The dynamics involved in  complex networks has become an
 important aspect of the  complex network studies in recent times. This
 problem includes, for example, the spreading of virus in the internet,
 synchronization of neurons in a brain, change of populations in a food
 web.
  Recently,  Pastor-Satorras  and Vespignani obtained unexpected results
 in this regard \cite{SIS0}.
 They studied the spread  of viruses  in complex networks and
 found that  no threshold of infection rate exists for the 
 susceptible-infected-susceptible model in the   random scale-free
 network with $\gamma \le 3$, if the size of the network $N$ is infinite.
 Though real networks such as the internet are finite-size network, this
 result implies that a virus with a small infection rate can spread over the
 whole network.
 We had previously presented another remarkable example of the unusual
 dynamics involved in complex networks\cite{ichinomiya, ichinomiya2}. We
 studied the Kuramoto synchronization
 in a random network of oscillators and found that the critical coupling
 for  synchronization
 becomes zero  in scale-free network with $\gamma \le 3$.  

 In these studies, the  mean-field approximation plays an essential role. For
 the mean-field approximation, we consider a model in which a  node $i$
 couples to another node $j$ with a strength proportional to ``mean
 coupling probability'' $ k_i k_j/k_{tot.}$, where
 $k_i$ and $k_{tot.}$ are  the degree of node $i$ and total number of
 edges, respectively. The dynamics in complex
 network is much simplified by this  approximation, and we can obtain
 analytical results. 
 However, this model differs from the
 original network model, in which each node couples to a finite number of
 nodes. It is remarkable that the mean-field approximation is in  good
 agreement with the numerical simulation result of a random network model.

 One of the objectives of this paper is to provide a 
  sound explanation for the mean-field  approximation. It is 
 unclear why the mean-field approximation performs  well in the random network
 model. The validity of the mean-field approximation, particularly with
  regard to the Kuramoto transition, is debatable. Moreno, Pacheko and
 Vazquez-Prada carried out numerical simulations on the Kuramoto transition in
 the Barab\'{a}shi-Albert network.\cite{Moreno1, Moreno2}. They concluded
  that the critical coupling $K_c$ is not 0 even if $N\rightarrow \infty$.
 Their conclusion seems to contradict the result of  mean-field
  theory  $K_c =0$, though this
  discrepancy is possibly due to the difference of the order parameter
  used in these papers. 
 Restrepo, Hunt, and Ott suggested that the
 the argument based on the largest eigenvalue of the network matrix is
 superior to that based on the  mean-field theory\cite{Restrepo}. They
 demonstrated that the
 mean-field approximation is valid for Erd\"{o}s-R\'enyi networks and random
 scale-free networks with $\gamma=3$, while this approximation does not
  hold in the case of scale-free networks with $\gamma=2$.
 However, they did not provide any explanation as to why  the mean-field
 theory works well  in some random network models. 
 An appropriate explanation to this question is a matter of great
  interest and significance.  


 The second objective  of this study is to extend the mean-field theory.
 Although the mean-field theory displays  good qualitative coincidence with
  numerical simulation, it is impossible to examine the  fluctuation
 of the variables by the mean-field approximation. Moereover, as
 noticed, we cannot apply the mean-field approximation in some network
 models. Therefore it is meaningful to make an approximation that
 covers a wider range of complex networks.
 In this paper, we
 demonstrate  that the distribution of  variables in the sparse random
 network model can be  
approximated by that  obtained from a globally coupled network, in which
 the distribution of the interaction between the nodes  is  given by a
 Gaussian random number. This result indicates that the dynamics in
 random network can be approximated more precisely by appropriate
 methods  such as dynamical mean-field theory\cite{Hertz}. 

 In order to realize the above-mentioned objectives,  we utilize  the
 path-integral approach. The
 path integral, which was originally  developed for application in quantum
 mechanics\cite{Feynman}, has also been applied to random impurity
 problems\cite{Martin,Dominicis}, random spin glasses\cite{Sommers,
 Sompolinsky, Crisanti}, neural networks\cite{Kree, Balak}, and
 oscillator
 systems\cite{Stiller}.  
 One of the advantages of this approach is that
 the average over  an ensemble of networks can be calculated easily.
 Limitations of   the path-integral include an  infinite number of integrals
 and obtaining a precise average over the ensemble, which is not usually
 possible. However, this  method enables  approximation of 
 the distribution of the variables  in a systematic manner. In particular,
  the mean-field  approximation can be derived as  the lowest order
 approximation of the path integral. The  methods  used by us 
 are  similar to that used by 
 Theumann for the Hopfield network model\cite{Theumann}. 
 
 The outline of this paper is as follows. In Sec. \ref{path-integral}, we
 present the general description of the dynamics of a  network model based on
 the   path-integral approach. We derive a formula
 that is general and can be applied to
 any network model in this section. In Sec. \ref{approximation} we
 present two
 approximations of the path-integral formula, mean-field approximation
 and perturbation. We also prove  that the dynamics of a  random
 network is essentially identical to that of a random Gaussian network. 
 In Sec.
 \ref{simulation}, we apply the analysis to  the Kuramoto transition in
 a random
 sparse network. We present the results of numerical simulation, which
 are  consistent with that obtained from the  analysis. To conclude, we
 discuss our obtained results.

\section{Path-integral approach to the dynamics of  a network model\label{path-integral}}
 In this section, we introduce the formalism to study the dynamics of a 
 network model using the path-integral approach. 
We consider the following differential equations for the  network model:
\begin{equation}
 \dot{x_i} = f_i(x_i) + \sum_{j=1}^N a_{i,j}g(x_i,x_j)+\xi_i(t),\label{network-equation}
\end{equation}
where $\xi_i(t)$ is  a  random force that satisfies
 $\langle \xi_i(t)=0\rangle$, $\langle\xi_i(t)\xi_j(t^{'}) \rangle =
\delta_{i,j}\delta(t-t^{'})\sigma^2$.
We assume $x_i =x_{i,0}$ at $t=0$.  
 In order to discuss the dynamics of this system, it is useful to introduce
 Matrin-Siggia-Rose(MSR) generating functional $Z$, which is  defined as\cite{Martin,Dominicis}
\begin{equation}
 Z[\{l_{i,k}\},\{\bar{l}_{i,k}\}] =\left(\frac{1}{\pi}\right)^{N N_t}\left\langle \int \prod_{i=1}^{N}
				    \prod_{k=0}^{N_t}d x_{i,k}d\bar{x}_{i,k}e^{-S}
  \exp(l_{i,k}x_{i,k}+\bar l_{i,k} \bar{x}_{i,k})J\right\rangle,
\end{equation}
where the action $S$ is given by
\begin{equation}
 S = \sum_{i,k}\left[\frac{\sigma^2 \Delta t}{2} \bar{x}_{i,k}^2 +   i
 \bar{x}_{i,k}\left\{x_{i,k}-x_{i,k-1}-\Delta t(f_i(x_{i,k-1})+
  \sum_j a_{i,j}g(x_{i,k-1},x_{j,k-1})) \right\}\right],
\end{equation}
 and $\langle \cdots \rangle$ represents  the average over the ensemble of
 networks.  $J$ is  the functional Jacobian term,
\begin{equation}
 J=\exp\left(-\frac{\Delta t}{2} \sum_{i,j,k} \frac{\partial (f_i(x_{i,k})+a_{i,j}
 g(x_{i,k},x_{j,k}))}{\partial x_{i,k}}\right). 
\end{equation}
 Though this term is necessary for
 the renormalization $Z(0)=1$, it is a little cumbersome to treat it
 in a practical calculation, such as the mean-field approximation or
 a perturbation.   Here we note that, as De Dominicis
 showed\cite{Dominicis}, the only effect of this Jacobian term is to
 subtract the nonretarded correlation function
 $\langle \bar{x}_{i,k} x_{j,k+k^{\prime}} \rangle$, where $k^{\prime}\ge 1$.
 In the following discussion, we omit this Jacobian term,
 remembering that we only consider the retarded correlation function. 
Maintaining $\Delta t N_t$ constant at the limit $\Delta t \rightarrow 0$, we
 obtain the MSR generating functional. 

We consider the network described by
\begin{equation}
 a_{i,j}=\left\{
\begin{array}{ll}
 1 & \textrm{ with probability $p_{i,j}$,}\\
 0 & \textrm {with probability $1-p_{i,j}$}.
\end{array}
\right.\label{networks}
\end{equation}
We note that $p_{i,j}$ can be a function of  variables such as $i$
  or $j$. For
  example, in the one-dimensional chain model,  $p_{i,j}$ is 1 if $|i-j| =1$, else it
  is 0.
 The average over all networks can be expressed as
\begin{equation}
 \left\langle\exp\left[\sum_{i,k}   i \Delta t \bar{x}_{i,k}\sum_j a_{i,j}
	     g(x_{i,k-1},x_{j,k-1})\right]\right\rangle =
 \prod_{i,j}\left[p_{i,j}\exp\left\{\sum_k  i \Delta t
			      \bar{x}_{i,k}g(x_{i,k-1},x_{j,k-1})\right\} + 1-p_{i,j}\right],\label{randomnet-eq1}
\end{equation}
 and we obtain
\begin{equation}
 \langle e^{-S} \rangle = \exp(-S_0) \prod_{i,j}\left[p_{i,j}\exp\left\{\sum_k i \Delta t
			      \bar{x}_{i,k}g(x_{i,k-1},x_{j,k-1})\right\} + 1-p_{i,j}\right],
\end{equation} 
where
\begin{equation}
 S_0= \sum_{i,k}\frac{\sigma^2 \Delta t}{2}\bar{x}_{i,k}^2 +  i
 \bar{x}_{i,k}\{x_{i,k}-x_{i,k-1}-\Delta t f_i(x_{i,k-1})
 \}.
\end{equation}
 The  above-mentioned expression  is a general one and can be applied to
 the  dynamics of
 any network model. However, it is often impossible to calculate
 the precise value of $\langle e^{-S} \rangle$, particularly in the case
 of nonlinear
 dynamics. We need an approximation to obtain the value of 
$\langle e^{-S} \rangle$. 
 In the next section, we approximate Eq. (\ref{randomnet-eq1}) by
 assuming $p_{i,k} \ll 1$ and $p_{i,j}p_{k,l} \ll p_{i,j}$ for any
 $i,j,k,$ and $l$.

\section{Approximation  of the MSR generating functional in
 a sparse random network model}\label{approximation}

 In this section, we develop an approximation for the MSR generating functional
 $Z$ in a sparse random network. For this, we assume $p_{i,j} \ll 1$  and
 $p_{i,j}p_{k,l}\ll p_{i,j}$  for any $i, j, k,$ and
 $l$. In the case of the Erd\"{o}s-R\'enyi model, $p_{i,j}$ is independent
 of $i$ and $j$; $p_{i,j}=\langle k \rangle /N$. Therefore 
 this assumption is valid for a sparse Erd\"{o}s-R\'enyi model, because
 $p_{i,j} p_{k,l}=\langle k \rangle^2/N^2 \ll p_{i,j}$. In the case of
 random network with distribution $P(k)$, we construct the network as
 follows. First, we define the ``degee'' of node $i$ as $k^{\prime}_i$,
 whose distribution concides with $P(k)$. Second, we connect the
 nodes $i$ and $j$ with probability
 $p_{i,j}=k^{\prime}_i k^{\prime}_j / \sum_i k^{\prime}_i$. Using  this
 procedure, we obtain the random network whose degree distribution is
 approximately given by $P(k)$. In this case, if the maximum degree of
 a node $k_{max}$ is much smaller than $N$, $k_{max} \ll N$,
 the assumption is satisfied. On
 the other hand, this assumption is not satisfied in the Watts-Strogatz
 model, because $p_{i,i+1} \sim 1$.

Since
 $p_{i,j} \ll 1$, the approximate value of the logarithm of the
 right-hand side  of
 Eq. (\ref{randomnet-eq1}) is expressed as follows:

\begin{eqnarray}
\ln\left( \prod_{i,j}\left[p_{i,j}\exp\left\{\sum_k  i \Delta t
			      \bar{x}_{i,k}g(x_{i,k-1},x_{j,k-1})\right\}
+ 1-p_{i,j}\right]\right)&\sim& \sum_{i,j} -p_{i,j}+p_{i,j}\exp\left\{\sum_k  i \Delta t
			      \bar{x}_{i,k}g(x_{i,k-1},x_{j,k-1})\right\}\nonumber\\
&=& \sum_{i,j}p_{i,j}\sum_{l=1}^{\infty}\frac{1}{l!}\left(\sum_k i
						     \Delta t \bar{x}_{i,k}g(x_{i,k-1},x_{j,k-1})\right)^l.\label{randomnet-eq2}
\end{eqnarray}

Therefore, we obtain
\begin{eqnarray}
 \langle e^{-S} \rangle &\sim& \exp\left[\sum_{i,k}\left\{\frac{-\sigma^2\Delta t}{2} \bar{x}_{i,k}^2 -  i
 \bar{x}_{i,k}\{x_{i,k}-x_{i,k-1}-\Delta t f_i(x_{i,k-1})
 \}\right\}\right]\nonumber\\
 &\times &
  \exp\left[\sum_{i,j}p_{i,j}\sum_{l=1}^{\infty}\frac{1}{l!}\left(\sum_k
  i\Delta t\bar{x}_{i,k}g(x_{i,k-1},x_{j,k-1})\right)^l\right]
\label{randomnet-eq3}.
\end{eqnarray}
 To calculate $Z$ from this equation, we need  an infinite number of 
integrals, and we cannot carry out
 this integration practically. However, this formula gives us  much
 information about the averaged dynamics of networks. In the following
 subsections, we consider two simple approximation schemes, the mean-field
 approximation and perturbation.

 \subsection{Mean-field approximation and beyond}
To begin with, we consider an approximation that ignores the  $l \ge 2$
 part of Eq.(\ref{randomnet-eq3}) and  obtain
\begin{equation}
 \langle e^{-S} \rangle \sim  \exp\left(\sum_{i,k}\left\{-\frac{\sigma^2 \Delta t}{2} \bar{x}_{i,k}^2 -  i
 \bar{x}_{i,k}\{x_{i,k}-x_{i,k-1}-\Delta t (f_i(x_{i,k-1})+\sum_j p_{i,j}g(x_{i,k-1},x_{j,k-1}))
 \}\right\}\right).\label{expand1}
\end{equation}

 This result demonstrates that the MSR generating functional for
 Eq.(\ref{network-equation}) can be approximated  as
 that  for the system described by
\begin{equation}
 \dot{x_i} = f_i(x_i) + \sum_j^N p_{i,j}g(x_i,x_j)+\xi_i(t).\label{global-eq1}
\end{equation}
This equation implies that the mean-field  approximation  neglects the
 contribution of the term  $l\ge 2$  in
Eq.(\ref{randomnet-eq3}). The mean-field approximation method is
based on two assumptions. First, the higher-order term in
$p_{i,j}$ in Eq.(\ref{randomnet-eq2}) is neglected, and, second, the
higher-order term in Eq. (\ref{randomnet-eq3}) is neglected. The former
 assumption
is  valid if  $p_{i,j} \ll 1$ for all values of $i$ and $j$. However, 
neglecting the higher order term is not always valid.
In order to examine this argument, we  study the effect of  the term
 $l=2$. From the Stratnovich-Hubbard transformation, we obtain the following:
\begin{equation}
 \exp\left[p_{i,j}\frac{1}{2}\left(\sum_k  i\Delta
			      t\bar{x}_{i,k}g(x_{i,k-1},x_{j,k-1})\right)^2\right]
 = \sqrt{\frac{1}{2 \pi p_{i,j}}}\int dr_{i,j}\exp\left[-\frac{r_{i,j}^2}{2 p_{i,j}}+ i \left(r_{i,j}\sum_k\Delta
			      t\bar{x}_{i,k}g(x_{i,k-1},x_{j,k-1})\right)\right].\label{expand2}
\end{equation} 
 By comparing Eqs. (\ref{expand1}), (\ref{global-eq1}), and
 (\ref{expand2}), we observe that  the MSR generating functional  is
 identical to  that
 of the system described by
\begin{equation}
 \dot{x_i} = f_i(x_i) + \sum_j^N (p_{i,j}+ r_{i,j}) g(x_i,x_j)+\xi_i(t),
\end{equation}
 where $r_{i,j}$ is a random number and  its distribution  is
 given by a Gaussian distribution, with a  mean value of 0 and a dispersion
 $\langle r_{i,j}^2 \rangle =p_{i,j}$.

 Sequential application of the
 Stratonovich-Hubbard transformation
 yields the contribution from the term $l=2^n$. For example, if  we
 consider the term $l=4$, then because
\begin{equation}
 \exp\left[p_{i,j}\frac{1}{4!}\left(\sum_k  i\Delta
			      t\bar{x}_{i,k}g(x_{i,k-1},x_{j,k-1})\right)^4\right]
 = \sqrt{\frac{3 }{2 \pi p_{i,j}}}\int
 dr^{\prime}_{i,j}\exp\left[-\frac{3r^{\prime 2}_{i,j}}{
2 p_{i,j}}+ \frac{r^{\prime}_{i,j}}{2}\left(\sum_k   i \Delta t\bar{x}_{i,k}
g(x_{i,k-1},x_{j,k-1})\right)^2\right],\label{expand4}
\end{equation}
 the Gaussian fluctuation with dispersion $\sqrt{p_{i,j}/3}$ is added to $p_{i,j}$
 in Eq.(\ref{expand2}). Therefore, $\langle P \rangle$ is given by the
 solution of 
\begin{equation}
 \dot{x_i} = f_i(x_i) + \sum_j^N (p_{i,j}+ r_{i,j}) g(x_i,x_j)+\xi_i(t),
\end{equation}
 where the distribution of $r_{i,j}$ is given by a  Gaussian  with dispersion
 $\sigma_r=\sqrt{p_{i,j}+r^{\prime}_{i,j}}$, and 
  the distribution of $r^{\prime}_{i,j}$ is also Gaussian with
 dispersion $\sigma_r^{\prime}=\sqrt{p_{i,j}/3}$. By sequential
 application of this transformation, we can obtain the effect of the
 term $l=2^n$. However, as these correction terms are  small for large $l$, the
 Gaussian random network is a  good approximation of the random sparse network.

 The estimation that utilizes the Stratonovich-Hubbard transformation is
 very effective in mapping the dynamics of the  sparse network ensemble
 onto the dynamics of
 globally coupled networks. This method is very useful, especially in the
 case where  the dynamics of Gaussian random networks is well 
 known. However,
 the effectiveness of this transformation is
 limited, because it can only consider the terms  $l=2^n$. This method
 does not elaborate on the effect of the terms $l=3,5,6,\cdots$. In
 the next
 subsection we realize that   the
 correction in  $Z$ resulting from the term  $l= 2 m +1$ is of the order
 $p_{i,j}^2$ in the random network.

\subsection{Perturbation}

 As shown in the preceding section, though the mean-field approximation and the
 Stratonovich-Hubbard transformation are very effective methods, they
 only consider a limited number of   terms. To examine the effect of
 other tems,  we use the  perturbation technique  for a network model in
 this section.
 The perturbation gives  a formal  estimate of the value
 of $\langle e^{-S} \rangle$. Furthermore,
 we note that this method is highly effective  because  it allows us to 
 estimate the accuracy of the approximation by an order of $p_{i,j}$.
 However, this method  has two drawbacks. First, it is typically
  impossible to  obtain the value of 
 $\langle e^{-S} \rangle$ in nonlinear physics. It is often difficult to
 obtain the value of $\langle e^{-S} \rangle$ even if there is
 no  interaction, and
 the perturbation can therefore only be applied to limited
 systems. Second,  bifurcation or phase transition cannot be obtained 
without including the 
 infinite order of perturbation in $p_{i,j}$. In general, the
 MSR generating functional $Z$ becomes singular at the bifurcation point. However,
 the finite order perturbation yields $Z$, which is a nonsingular function
 of $p_{i,j}$. Therefore, it is impossible to examine a phase
 transition by perturbation.
 However, perturbation often yields 
 important information.
 In this section, we demonstrate that   the correction from  the odd $l$
 term is of the order  $p_{i,j}^2$.

 We begin with Eq. (\ref{randomnet-eq3}). On expanding $\exp(\sum
 p_{i,j}\cdots)$, we  obtain
\begin{eqnarray}
  \langle e^{-S} \rangle &=& \exp\left[\sum_{i,k}\left\{-\frac{\sigma^2\Delta t}{2} \bar{x}_{i,k}^2 -  i
 \bar{x}_{i,k}\{x_{i,k}-x_{i,k-1}-\Delta t f_i(x_{i,k-1})
 \}\right\}\right]\nonumber\\
 &\times & \sum_{m=0}^{\infty} \frac{1}{m!}\left[\sum_{i,j}p_{i,j}\sum_{l=1}^{\infty}\frac{1}{l!}\left(\sum_k i\Delta t\bar{x}_{i,k}g(x_{i,k-1},x_{j,k-1})\right)^l\right]^m.
\end{eqnarray}
 We first consider  the correction from  the term $l=3$. Since
 the effect of the term $l=2$  can be expressed  as the fluctuation in
 $p_{i,j}$, we consider 
\begin{eqnarray}
  e^{-S}  &\sim&  \exp\left(\sum_{i,k}\left\{-\frac{\sigma^2 \Delta t}{2} \bar{x}_{i,k}^2 -  i
 \bar{x}_{i,k}[x_{i,k}-x_{i,k-1}-\Delta t (f_i(x_{i,k-1})+\sum_j p^{'}_{i,j}g(x_{i,k-1},x_{j,k-1}))
 ]\right\}\right.\nonumber\\
&+&\left.\sum_{i,j}\frac{p_{i,j}}{3!}\left(\sum_k  i \Delta t
    \bar{x}_{i,k}g(x_{i,k-1},x_{j,k-1})\right)^3\right)\label{expand3-0}
\end{eqnarray}
 To account for the effect of the term $l=3$, we treat this term
 using  perturbation. We expand the term $l=3$ as 
\begin{eqnarray}
 \exp\left[\sum_{i,j}\frac{p_{i,j}}{3!}\left(\sum_k  i \Delta t
					\bar{x}_{i,k}g(x_{i,k-1},x_{j,k})\right)^3\right] & = & 1+ \sum_{i,j}\frac{p_{i,j}}{3!}\left(\sum_k  i \Delta t
      \bar{x}_{i,k}g(x_{i,k-1},x_{j,k-1})\right)^3\nonumber\\
&+&\frac{1}{2}\left\{\sum_{i,j}\frac{p_{i,j}}{3!}\left(\sum_k  i \Delta t\bar{x}_{i,k}g(x_{i,k-1},x_{j,k-1})\right)^3\right\}^2+\cdots.\label{expand3-1}
\end{eqnarray}
 Here we assume $p_{i,j} p_{k,l} \ll p_{i,j}$ for any value of $i,j,k,$ and
 $l$ again.

 Based on this assumption, $\langle e^{-S} \rangle $ can be  approximated as
\begin{equation}
 \langle e^{-S} \rangle =\left\{1+ \sum_{i,j}\frac{p_{i,j}}{3!}\left(\sum_k  i \Delta t
      \bar{x}_{i,k}g(x_{i,k-1},x_{j,k-1})\right)^3\right\} e^{-S_1},\label{expand3-basic}
\end{equation}
where
\begin{equation}
 S_1 = \sum_{i,k}\left\{\frac{\sigma^2 \Delta t}{2} \bar{x}_{i,k}^2 +  i
 \bar{x}_{i,k}[x_{i,k}-x_{i,k-1}-\Delta t (f_i(x_{i,k-1})+\sum_j p^{'}_{i,j}g(x_{i,k-1},x_{j,k-1}))
 ]\right\}
\end{equation}
 To calculate the contributions to $Z$ from these terms, it is convenient to define
\begin{eqnarray}
  P(\{\epsilon_{i,k}\}) &=& \left(\frac{1}{\pi}\right)^{\frac{N N_t}{2}}\int
 \prod_{i,k}d\bar{x}_{i,k}\exp\left(\sum_{i,k}\left\{-\frac{\sigma^2 \Delta t}{2} \bar{x}_{i,k}^2 -  i
 \bar{x}_{i,k}[x_{i,k}-x_{i,k-1}-\Delta t(f_i(x_{i,k-1})\right.\right.\nonumber\\
&&+\left. \left. 
  \sum_j p^{\prime}_{i,j}g(x_{i,k-1},x_{j,k-1})-\epsilon_{i,k}) ]\right\}\right).\label{define-mother-functional}
\end{eqnarray}
 Since
\begin{equation}
 \int \prod_{i,k}d {\bar x}_{i,k}(- i \Delta t)^3 {\bar x}_{i_1,k_1}{\bar
  x}_{i_1,k_2}{\bar x}_{i_1,k_3} e^{-S_1}=
  \frac{\partial^3}{\partial \epsilon_{i_1,k_1}\partial
  \epsilon_{i_1,k_2}\partial \epsilon_{i_1,k_3}}P(\{\epsilon_{i,k}\})|_{\epsilon_{i,k}\rightarrow 0},\label{cont-mother-functional}
\end{equation}
 the MSR generating functional $Z$ can be calculated from Eq.
 (\ref{expand3-basic}) if the  differential of $P$ is known. 
We first consider the differential of $P$ in the case where $k_1, k_2$,
 and $k_3$ are distinct. 
On integrating Eq. (\ref{define-mother-functional}), we obtain
\begin{equation}
 P(\{\epsilon_{i,k}\}) = C \exp\left[-\sum_{i,k}\frac{1}{2 \sigma^2\Delta
			     t}\left\{x_{i,k} -x_{i,k-1} - \Delta t
				\left(f_i(x_{i,k-1}) + \sum_j
				 p^{\prime}_{i,j} g(x_{i,k-1},x_{j,k-1})-\epsilon_{i,k}\right)\right\}^2\right],\label{expand3-2}
\end{equation}
 where $C=({1}/{\sigma^2 \Delta t})^{NN_t/2}$. We note
 that the integration of $P(\{\epsilon_{i,k}\})$ over $x_{i,k}$ is
 $O(1)$, while $P(\{\epsilon_{i,k}\})$ is $O((\Delta t)^{3/2})$.
On differentiating  Eq. (\ref{expand3-2}) with respect to
 $\epsilon_{i_1,k_1}$,
 we obtain
\begin{equation}
 \frac{\partial P(\{\epsilon_{i,k}\})}{\partial
  \epsilon_{i_1,k_1}}\Bigr|_{\epsilon = 0}=\frac{C}{\sigma^2}\left\{x_{i_1,k_1} -x_{i_1,k_1-1} - \Delta t
				\left(f_{i_1}(x_{i_1,k_1-1}) + \sum_j
				 p^{\prime}_{i_1,j} g(x_{i_1,k_1-1},x_{j,k_1-1})\right)\right\}e^{-S_1^{\prime}},\label{expand3-3}
\end{equation} 
where
\begin{equation}
 S_1^{\prime} = \sum_{i,k}\frac{1}{2 \sigma^2\Delta
			     t}\left\{x_{i,k} -x_{i,k-1} - \Delta t
				\left(f_i(x_{i,k-1}) + \sum_j
				 p^{\prime}_{i,j} g(x_{i,k-1},x_{j,k-1})\right)\right\}^2.\label{s1prime}
\end{equation}
 On differentiating  Eq. (\ref{expand3-3}) with respect to
 $\epsilon_{i_1,k_2}$ and
 $\epsilon_{i_1,k_3}$ we obtain $\partial^3 P/\partial
 \epsilon_{i_1,k_1}\partial \epsilon_{i_1,k_2}\partial
 \epsilon_{i_1,k_3}$. 
 However, we temporarily consider the term 
$\partial P/\partial \epsilon_{i,k}$,
 because further differentiations with respect to $\epsilon_{i_2,k_2}$
 and $\epsilon_{i_3,k_3}$ do not modify the following discussion. 

 In the limit $\Delta t \rightarrow 0$, $\exp(-S^{\prime}_1)$ approaches
 to  the $\delta$ function
 $\delta(x_{i,k} -x_{i,k-1} - \Delta t	[f_i(x_{i,k-1}) + \sum_j
 p^{\prime}_{i,j} g(x_{i,k-1},x_{j,k-1})])$. Therefore, in the limit 
 $\Delta t \rightarrow 0$, Eq. (\ref{expand3-3}) always attains the
 value 0.
 However,  $\Delta t N_t$ is maintained constant when  the limit $\Delta t
 \rightarrow 0$ is taken.
  Therefore, the sum of the  integrals of $\partial P/\partial \epsilon_{i,k}$
 over $x_{i,k}$  may have a  finite value at the limit
  $\Delta t \rightarrow 0$, if Eq.(\ref{expand3-3}) has a magnitude
 $O(\Delta t)$. 
 In order to obtain a more accurate estimate of  Eq.(\ref{expand3-3}),
 we express $S^{\prime}_1$  as 
\begin{eqnarray}
 S_1^{\prime} &=& \sum_{i,k}\frac{1}{2 \sigma^2\Delta
			     t}\left\{x_{i,k} -x_{i,k-1} - \Delta t
				\left(f_i(x_{i,k-1}) + \sum_j
				 p^{\prime}_{i,j}
				 g(x_{i,k-1},x_{j,k-1})\right)\right\}^2 \nonumber\\
&+& \frac{1}{\sigma^2} \left(f_i(x_{i,k}) +\sum_j
       p^{\prime}_{i,j}g(x_{i,k},x_{j,k})\right)(x_{i,k+1}-x_{i,k})+
O(\Delta t) +(\mbox{the terms that do not include  $x_{i,k}$}).\label{expand3-6}
\end{eqnarray}
We consider the integral $\int d x_{i,k} h(x_{i,k}) \partial
 P/\partial \epsilon_{i,k}$, where $h(x_{i,k})$
 is an arbitrary nonsingular function. Since
 the value of  $h(x_{i,k})[x_{i,k}-x_{i,k-1}-\Delta t \{f_{i}(x_{i,k-1}) +
 p^{\prime}_{i,j} g(x_{i,k-1},x_{j,k-1})\}]$ is 0 at
 $x_{i,k}=x_{i,k-1}+\Delta t (f_{i}(x_{i,k-1}) + p^{\prime}_{i,j}g(x_{i,k-1},x_{j,k-1}))$, we
 introduce $y_{i,k}=x_{i,k}-x_{i,k-1}-\Delta t \{f_{i}(x_{i,k-1}) + p^{\prime}_{i,j}g(x_{i,k-1},x_{j,k-1})\}$,
 and expand using the Taylor expansion.
\begin{equation}
 h(x_{i,k})\exp\left\{ \frac{1}{\sigma^2} \left(f_i(x_{i,k}) +\sum_j
       p^{\prime}_{i,j}g(x_{i,k},x_{j,k})\right)(x_{i,k+1}-x_{i,k})+
O(\Delta t) +\cdots\right\} = \sum_{m=1}^\infty h_m y^m_{i,k}.
\end{equation}
Therefore, we obtain
\begin{eqnarray}
 \int d x_{i,k} h(x_{i,k}) \frac{\partial P}{\partial
  \epsilon_{i,k}}|_{\epsilon =0 }& =&
  C \int dy_{i,k} \sum_m h_m y^m_{i,k} \exp(-y^2_{i,k}/2 \sigma^2 \Delta t)\nonumber\\
 &=& C \int dy^{\prime}_{i,k}\sum_m (2 \sigma^2 \Delta t)^{{m+1}/{2}}
  h_m y^{\prime m}_{i,k}
  \exp(-y^{\prime 2 }_{i,k}).
\end{eqnarray}
 Since the value of this integral becomes 0 if $m$ is odd, the leading
 order of this integral is obtained from the term  $m=2$, and has a
 magnitude of  $O(\Delta t^{3/2})$. Since the magnitude of this
 contribution is smaller than $O(\Delta t)$, we
 can neglect  this term at the limit $\Delta t \rightarrow 0$. 

 Therefore, if  $k_1$, $k_2$, and $k_3$  are unequal, then the value of
 the integral of the second term in Eq. (\ref{expand3-basic})
 becomes  0. In the
 case where  $k_1=k_2=k_3$,  $\partial^3
 P(\epsilon)/\partial \epsilon^3$ is negligible as  $\Delta t \rightarrow
 0$ from a similar argument. In general,  the contribution from the term
 $\partial^m P(\epsilon)/\partial \epsilon^m$ is negligible for odd $m$,
 because it is expressed in the form
 $e^{-S_1^{\prime}}(x_{i,k} -x_{i,k-1} - \Delta t
(f_i(x_{i,k-1}) + \sum_j p^{\prime}_{i,j} g(x_{i,k-1},x_{j,k-1})))
 \times(\mbox{nonsingular function})$. 

 Similarly,
 we can prove  that  $[{\partial^m}/{\partial
 \epsilon_{i_1,k_1}\cdots \partial \epsilon_{i_m,k_m}}]P$ is nonzero if
 and only if each $\epsilon_{i,k}$ appears $2n$ times in the delimiter. 
 From these results, we conclude that the contribution of 
 the term $l=3$ to $Z$ is of the order $p_{i,j}^2$. In addition, we
 conclude that the contribution
 from the term $l= 2 m + 1$ is of  the order  $p_{i,j}^2$.
 The contribution from the term $l=2^m (2 m^{\prime}+1)$ is  also
 estimated  by the Stratonovich-Hubbard transformation. In the case of
 the 
 random network model, the correction due to these terms is small. 
 For example, the contribution  from the
 term $l=6$ to the MSR generating functional
 term is estimated using 
\begin{equation}
 \exp\left[p_{i,j}\frac{1}{6!}\left(\sum_k i \Delta t \bar{x}_{i,k}
			       g(x_{i,k-1},x_{j,k-1})\right)^6
     \right]=\sqrt{\frac{180}{\pi p_{i,j}}}\int dr_{i,j}
 \exp\left[-\frac{r^2_{i,j}}{ p_{i,j}}+ i r_{i,j}\left(\sum_k i \Delta t
  \bar{x}_{i,k} g(x_{i,k-1},x_{j,k-1})\right)^3\right].
\end{equation}
 The dispersion of $r_{i,j}$ is $\sqrt{p_{i,j}/90}$ and the contribution
 of this term is much smaller compared to  that from the Gaussian fluctuation
 obtained from the term $l=2$.

 From the discussion based on the Stratonovich-Hubbard transformation and
 perturbation, we demonstrated  that the MSR generating functional  for
 the dynamics of a  random sparse
 network model is almost identical to that for the dynamics of a random
 Gaussian network. In the next section, we demonstrate  that the above
 analysis 
 is consistent with the result of a  numerical simulation of the
 Kuramoto transition
 in a network model.

\section{Example: the Kuramoto transition}\label{simulation}

 In the previous section, we developed a general scheme to approximate
 the dynamics of the random sparse network and found its  dynamics can be 
 described by
\begin{equation}
 \dot{x_i} = f_i(x_i) + \sum_j^N (p_{i,j}+ r_{i,j}) g(x_i,x_j)+\xi_i(t) 
\end{equation}
 when $p_{i,j} \ll 1$. In this case, the distribution of
 $r_{i,j}$ is provided by a Gaussian with dispersion
 $\sigma = \sqrt{p_{i,j}}$. In this
 section, we apply this approximation to the  dynamics of oscillators
 in random networks.

  We consider a random network
 of oscillators
\begin{equation}
 d\theta_i/dt = \omega_i + K\sum_j a_{i,j}\sin(\theta_j-\theta_i)
\end{equation}
  where $\theta_i$ and  $\omega_i$ represent  the phase and velocity of
 the oscillator  $i$, respectively. The value of $a_{i,j}$ is 1 if nodes
 $i$ and $j$ are connected, otherwise it is 0. We note that the  random
  Gaussian matrix needs to be considered as symmetric. We consider
  the case where  the
 distribution of
  $\omega_i$ is given by $g(\omega)=({1}/{\sqrt{2 \pi
  \sigma_\omega}})\exp((\omega-\omega_0)^2/2 \sigma_{\omega}^2)$,
  and $a_{i,j}$ represents a random network with its mean degree
 $k_0$.
 The  above discussion suggests that the dynamics of this network can be
  approximated using  the following equation:
\begin{equation}
 d\theta_i/dt =\omega_i +\sum_j\left(\frac{k_0}{N}+r_{i,j}\right)K\sin(\theta_j-\theta_i),
\end{equation} 
where the distribution of $r_{i,j}$ is given by
$P(r_{i,j}) = \sqrt{{N}{2\pi k_0}}\exp(- N r_{i,j}^2/(2k_0))$.
This model is similar to the dynamic glass
model proposed by Daido\cite{Daido}. However, the mean
interaction between oscillators is positive in our model, while it is 0
in  Daido's model. 

 It is difficult to calculate analytically the dynamics of this
 globally coupled model.
 In this section, we present the numerical
 results for the random sparse and random Gaussian networks. For this
 simulation, we set $N=1000$, $2 \sigma_{\omega}^2= 1.0$, $\omega_0 =0$, and
 $k_0 = 10$. 
 The result obtained is  averaged over 50 different networks.

\begin{figure}[t]
 \resizebox{.4\textwidth}{!}{\includegraphics{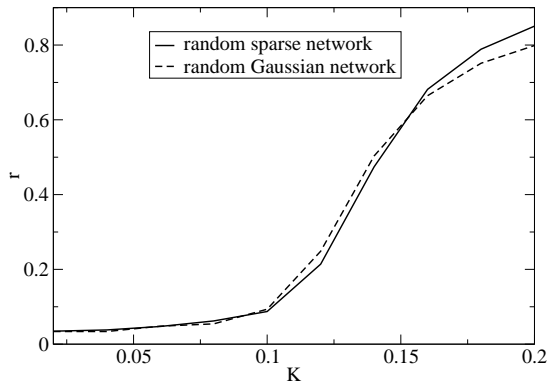}}
 \caption{The coupling dependence of the order parameter for random sparse
  and random Gaussian networks.\label{figure-order-parameter}}
\end{figure}

 First we examine the coupling dependence of the order parameter $r$. In
 our previous paper, we defined the order parameter $r$ as $r=\sum_i k_i
 e^{i\theta_i}/\sum k_i$ for a random sparse network model. However, it is
 difficult to define such an order parameter for a  random Gaussian
 model. In this paper, we therefore use
 $\langle e^{i\theta} \rangle $  as the order parameter for a random
 Gaussian network, and
  $r=\sum_i k_i e^{i\theta_i}/\sum k_i$  for a random sparse network model.
 Although these two order parameters are distinct, the difference
 between them is small, because  the distribution of degree has a strong
 peak at $k=k_0$ for a random sparse network. 
 The values of $r$ are plotted in Fig. \ref{figure-order-parameter} for 
both the networks for the range $K=0.02$-0.20. In both
 these  models, the order parameter remains almost constant for $K$
 values less than 0.1. There is a rapid increase in $r$ for $K$ values
 greater than 0.1. The values of
 $r$ coincide qualitatively for these two models.
 When $K=K_c$, a sharp transition, given by $r\propto \sqrt{K-K_c}$, is
 observed in the mean-field approximation. This sharp transition gets
 smeared out in a random sparse network. The Gaussian
 model approximates this smearing well.

\begin{figure}[t]
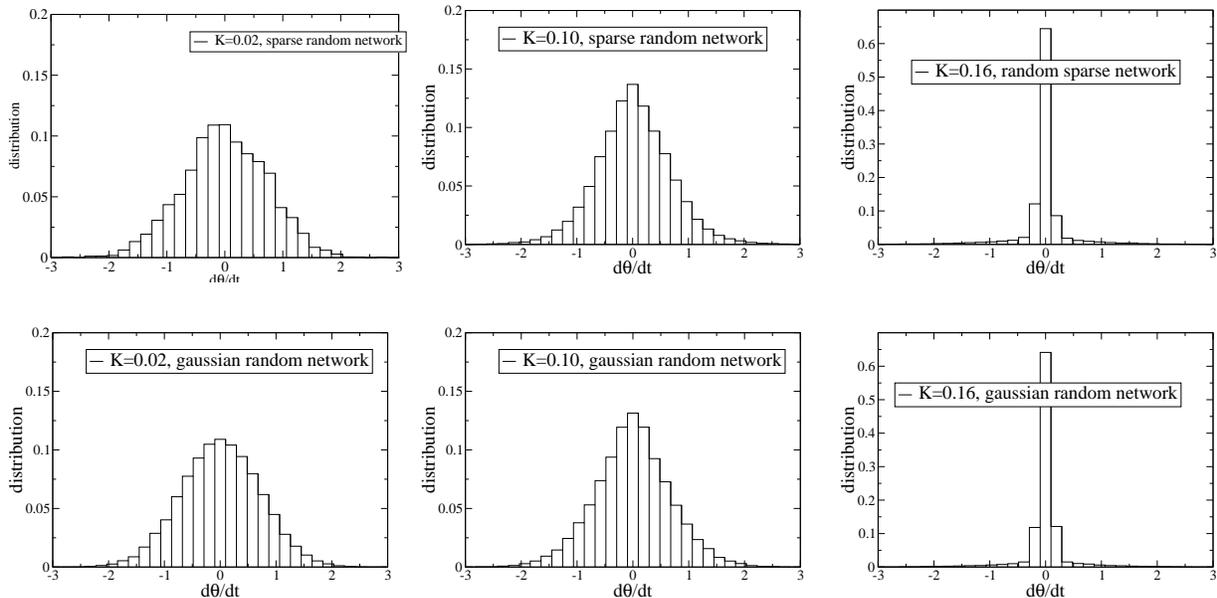

  \resizebox{.3\textwidth}{!}{\includegraphics{figure2a.ps}}
  \resizebox{.3\textwidth}{!}{\includegraphics{figure2b.ps}}
  \resizebox{.3\textwidth}{!}{\includegraphics{figure2c.ps}}
  \resizebox{.3\textwidth}{!}{\includegraphics{figure2d.ps}}
  \resizebox{.3\textwidth}{!}{\includegraphics{figure2e.ps}}
  \resizebox{.3\textwidth}{!}{\includegraphics{figure2f.ps}}
 \caption{The distribution of $d\theta/dt$ at $K=0.02, 0.10$, and 0.16 for
 a random sparse network (upper) and a random Gaussian network (lower).\label{veldist}}
\end{figure}

 The order parameters being identical is not unusual, because their
 obtained values were  close to those  obtained using the  mean-field
 theory. We now explain the distribution of velocity
 $d \theta_i/d t$. In the mean-field approximation, $d\theta/d t $ has
 a $\delta$-function-like  peak
   at $d \theta/dt = 0$. However, if the coupling
 between the oscillators is  random, the strong peak at $d\theta/dt$ will get
 smeared. In Fig. \ref{veldist},  the distribution of $d\theta/dt$
 for sparse random and random Gaussian networks is plotted. At $K=0.02$,
 there is an absence of  synchronization and the distribution of
 $d \theta/ dt $ is Gaussian-like. On the other hand, the oscillators are well
 synchronized and the distribution has a strong peak at
 $d\theta /dt = 0$ when $K=0.16$.
 For the present study, we focus on the distribution at $K=0.10$. This
 value of $K$ is close to the critical point,  and we suggest that the large
 fluctuation appears at this point. In the case of sparse networks, the
 peak at $d\theta /dt =0$ is sharper at $k=0.10$ than at
 $k=0.02$. The same tendency is observed in the case of a Gaussian
 network.
 For example, we observe that $P(-0.1 < d\theta/dt <0.1)=0.137$
 for a sparse random network. This value  is close to 
 $P(-0.1 < d\theta/dt <0.1)=0.131$ obtained from a random Gaussian
 network.
 This
 consistency  in the observed value suggests that a random sparse
 network can be approximated by a  Gaussian random network.

\begin{figure}[t]
 \resizebox{.4\textwidth}{!}{\includegraphics{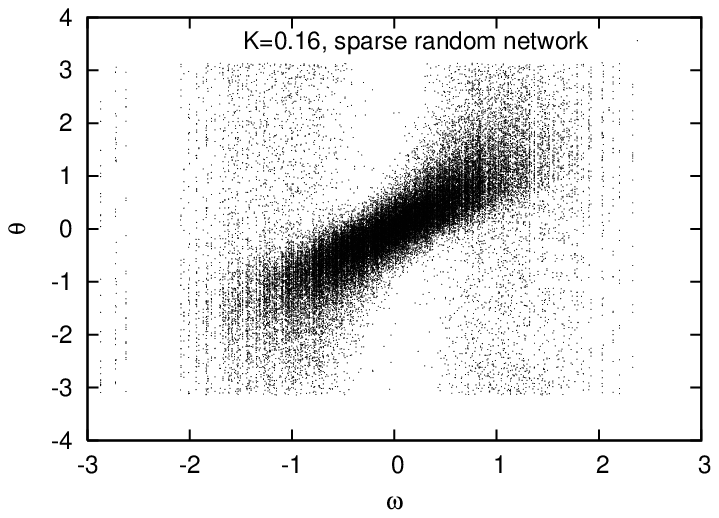}}
 \resizebox{.4\textwidth}{!}{\includegraphics{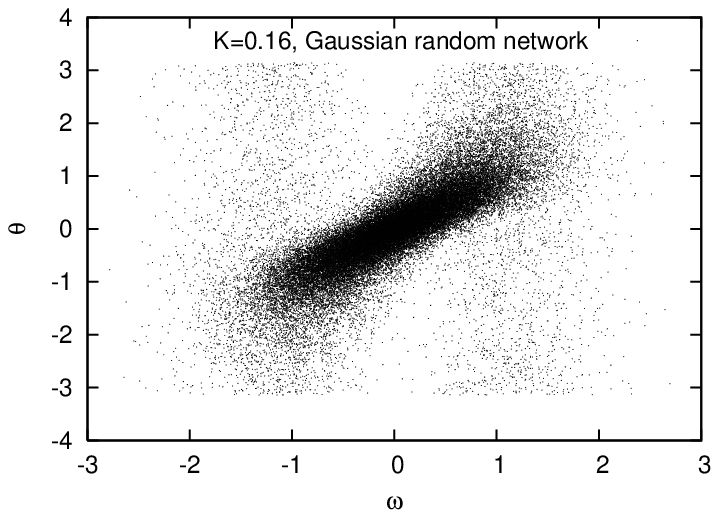}}
 \caption{Phase distribution in sparse and Gaussian random networks at
 $K=0.16$. \label{phasedist}}
\end{figure}

 Finally, we present the distribution of the phase $\theta$ for both
 networks. The phase distribution in the $(\omega, \theta)$ plane
  at $K=0.16$ is shown in Fig. \ref{phasedist}. Although the coupling
 strength is sufficiently large for synchronization, the phase 
 does not entirely lie  on a single line obtaind from  the mean-field
 approximation method, $\theta = \arcsin(\omega/Kk_0r)$. In order to
 observe  the dispersion
 around the mean-field line, we present the phase distribution of 
 oscillators with $|\omega| \le 0.05$ in Fig.\ref{phasedist0}. In this
 region, $|\arcsin(\omega/K k_0 r)|$ is less than 0.05 and  the
 $\omega$ dependence of the phase distribution can be neglected.  In
 both the models, the phase distribution lies  in a wide range of
 $\theta$. The dispersion $\sigma$ for these two figures is $\sigma^2
 = 0.105$ and $0.122$ for the random sparse and the random
 Gaussian network, respectively.
 Since these two values coincide qualitatively, 
 random Gaussian network is a good approximation of the random sparse
 network.  
\begin{figure}[t]
 \resizebox{.4\textwidth}{!}{\includegraphics{figure4a.ps}}
 \resizebox{.4\textwidth}{!}{\includegraphics{figure4b.ps}}
 \caption{Phase distribution in random sparse and random Gaussian
 networks at $K=0.16$. \label{phasedist0}}
\end{figure}

\section{Conclusion and Discussion}

 In this paper, we studied the dynamics  of a  random
 network model using  the path-integral approach. We identified that the
 mean-field approximation is the lowest-order approximation of
 $p_{i,j}$ and $l=1$, as shown in Eq. (\ref{expand1}). We also
 demonstrated  that
 the contribution of the term $l = 2^n$ can be described by the
 fluctuation
 of coupling in the globally coupled approximation method. The
 contribution of the odd $l$ terms is difficult to
 estimate, though it is of the order $p_{i,j}^2$. We applied these general
 results to the Kuramoto transition, and observed a  good agreement with
 numerical simulations. 

 The path-integral approach  developed through this study is a general
 one and is  applicable to dynamics of any  random network. In
 particular, if 
 the precise result for a randomly coupled  model is known, a good
 approximation can be obtained for  random sparse network models. There
 are several
 models, such as the replicator model\cite{Diederich}, for which the exact
 results are known for a Gaussian random
 network . Our analysis
 proves that the dynamics of random sparse networks can be easily obtained
 for such models.

 The analysis presented in this study is limited to the dynamics in a
 random network model. In the case of another network model, we 
 need to include the higher-order terms to evaluate the MSR generating
 functional.  It is usually difficult to carry out such a calculation.
 However, our result provide much informations regarding the validity
 of the mean-field approximation. For example, the mean-field
  approximation is applicable if  $p_{i,j} p_{k,l} \ll p_{i,j}\ll
 1$.
 On the other hand,
 such an approximation is not applicable to the dynamics of a  highly
 clustered network. In such a network,
  $p_{i,j}p_{j,k}p_{k,i}\sim O(p_{i,j}p_{j,k})$,
 $p_{i,j}p_{k,l} \ll p_{i,j}$ cannot be assumed and  the contribution
 from the neglected terms needs to be calculated.
 It is usually believed that the dynamics of networks
 with high clustering coefficients cannot be approximated using the
 mean-field approximation method because of the high clustering
 coefficient. However, our analysis reveals that the validity of the
 mean-field approximation methods depends on the value of
 $p_{i,j} p_{k,l}$ and
 $p_{i,j}$. For example, the mean-field approximation method cannot be
 applied to the
 square-lattice model even if the clustering coefficient is zero,
 because the value of $p_{i,j}p_{k,l}$ can be as large as that of $p_{i,j}$. 

 We also discuss other studies conducted on the  Kuramoto transition
 in random
 network models.  Restrepo {\it et al.} examined the  mean-field
 theory and studied the Kuramoto transition \cite{Restrepo}. They concluded that
  synchronization occurs when $K$ satisfies the relation
 $K > 2/\pi g(0) \lambda$, where $\lambda$ is the largest eigenvalue of 
the network matrix $a_{i,j}$ and $g(0)$ is the density of the oscillators at
 $\omega=0$. They stated that the mean-field
 approximation, which was developed by us  in  previous papers,
 functions  only when
 $r_i \propto k_i$, where $r_i$ is the local field defined as
 $r_{i}= \langle \sum_j a_{i,j} e^{i (\theta_i-\theta_j)} \rangle_t$,
 where $\langle \cdots \rangle_t$ means the average over a long time
 interval. However, they did not explain the reason why this
 assumption is valid  some random network models, though they stated
 that there exists some relationship between the eigenvectors of $a_{i,j}$
 and degree of each node. In this paper, we demonstrated that the mean-field
 theory is an approximation that considers  only the term $l=1$ in the
 MSR generating functional. In this case, the mean-field approximation
 coincides with the discussion obtained from the largest eigenvalues,
 because the largest eigenvalue of the matrix $p_{i,j}=x_i x_j$ is
 $\sum_i x_i^2$ and its eigenvector $v$ is given by
 $v=(x_1,x_2,\cdots x_n)$. In the random network model,
 $p_{i,j} = k_i k_j/N\langle k \rangle$, where $k_i$ and $k_j$ are
 the degrees of the nodes $i$ and $j$. Therefore, the largest eigenvalue of
 this matrix is $\langle k^2\rangle/\langle k \rangle$,
 and  the critical condition for
 synchronization in the mean-field approximation becomes identical to
 that  in the
 discussion based on eigenvalues. In order to examine the applicability of the
 mean-field approximation,  the term  $l=2$ should be considered. In the
 case of a random matrix,  the largest eigenvalue with a dispersion $p$
 is expressed by $2 \sqrt{ N p} = 2 \sqrt{ \langle k \rangle }$ based on 
 Wigner's semicircle law\cite{Wigner}.
 This result suggests that the mean-field approximation can be applied if 
  $\sqrt{\langle k \rangle } \ll \langle k^2 \rangle/\langle k \rangle
 $.
 In order to examine this, we consider the
 matrix $M+G$, where $M$ is the matrix obtained from the  mean-field
 approximation and
 $G$ the Gaussian random matrix, i.e., the distribution  of each
 element of the matrix is Gaussian with dispersion $\sqrt{k}$. As
 observed earlier, the largest eigenvector $v$
 of matrix $M$ satisfies the condition $M v = \lambda v$, where
 $\lambda=\langle k^2 \rangle/\langle k \rangle$. On the other hand,
 $|G v|$ is  of the order 
 $\sqrt{2 k}|v|$, because all eigenvalues of $G$ lie  between
 $-\sqrt{2 k}$ and $\sqrt{2 k}$. Therefore, $|(M+G)v|$ equals
 approximately  
 $\lambda |v|$ based on  the assumption that $\lambda \gg \sqrt{k}$, and
 the direction of $(M+G)v$ is approximately identical to $v$.  
 It should be  noted that all vectors $u$ that  are perpendicular to $v$,
 i.e., $(u,v)=0$, satisfy the condition $M u = 0$. This implies that 
 $|(M+G)u|\le \sqrt{2 k} |u| \ll \lambda |u|$.
 Therefore, the  largest
 eigenvalue and corresponding  eigenvector of $M+G$ can be approximated as
 $\langle k^2 \rangle /\langle k \rangle$ and $v=(k_1,\cdots, k_n)$, 
respectively. Therefore, the mean-field approximation is a suitable 
 approximation if $\langle k \rangle$ is sufficiently large.
 In the case of a scale-free network, the spectrum density differs
 from that suggested by Wigner's law and the above-mentioned conclusion
 should be modified. However,
 this discussion suggests that  the validity of the mean-field approximation is
 determined by the largest eigenvalues of the mean-field matrix $M$ and
 random matrix $G$. If the largest eigenvalue of  matrix $G$ is as
 large as $\lambda$, the mean-field approximation is not valid.   
 Based on the idea presented in this paper, the claim made by
  Restrepo {\it et al.} implies that the term $l=2$ must be included 
 in order to discuss the critical behavior more accurately, especially
 in the case
 of a scale-free network with $\gamma=2$.  Therefore, their work was 
 not a  denial of the mean-field theory, but an extension of it.


\begin{acknowledgements}
 We would like to acknowledge Y. Nishiura, Y. Kuramoto, M. Iima, and
 T. Yanagita for fruitful discussion. 
\end{acknowledgements}


\begin{thebibliography}{99}
 \bibitem{Watts} D. J. Watts and S. H. Strogatz, Nature(London){\bf 393},
	 440(1998). 
 \bibitem{Barabashi-Review} R. Albert and A. -L. Barab\'ashi,
	 Rev. Mod. Phys. {\bf 74}, 47(2002).
 \bibitem{Newman} M. J. E. Newman, SIAM Rev. {\bf 45},167 (2003).
 \bibitem{Dorogovstev} S. N. Dorogovtsev and J. F. F. Mendes,
	 {\it Evolution of Networks} ( Oxford University Press, London, 2003).  
 \bibitem{SIS0} R. Pastor-Satorras and A. Vespignani, Phys. Rev. Lett. {\bf 86}, 3200(2001). 
 \bibitem{ichinomiya} T. Ichinomiya, Phys. Rev. E {\bf 70},
	 026116(2004). 
 \bibitem{ichinomiya2} T. Ichinomiya, Prog. Theor. Phys. {\bf 113}, 1(2005).
 \bibitem{Moreno1} Y. Moreno and  A. F. Pacheco, Europhys. Lett. {\bf 68},
	 603(2004).
 \bibitem{Moreno2} Y. Moreno, M. Vazquez-Prada, and A. F. Pacheco,
	 Physica A{\bf 343} 279(2004).
 \bibitem{Restrepo} J. G. Restrepo, E. Ott and B. R. Hunt, Phys. Rev. E
	 {\bf 71} 036151(2005).
 \bibitem{Hertz} J. A. Hertz, J. Phys. C {\bf 16} 1219(1983).
 \bibitem{Feynman} R. P. Feynman and A. R. Hibbs, {\it Quantum
	 mechanics and path integrals}( McGraw-Hill, New York, 1965).
 \bibitem{Martin} P. C. Matrin, E. Siggia and H. Rose, Phys. Rev. A {\bf
	 8} 423(1973).
 \bibitem{Dominicis} C. De Dominicis, Phys. Rev. B {\bf 18}, 4913(1978).
 \bibitem{Sommers} H. J. Sommers, Phys. Rev. Lett. {\bf 58}, 1268(1987).
 \bibitem{Sompolinsky} H. Sompolinsky  and A. Zippelius, Phys. Rev. B
	 {\bf 25}, 6860 (1982).
 \bibitem{Crisanti} A. Crisanti  and H.Sompolinsky, Phys. Rev. A {\bf
	 36}, 4922 (1987).
 \bibitem{Kree} R. Kree and  A. Zippelius, Phys. Rev. A {\bf 36}, 4421
(1987). 
 \bibitem{Balak}  J. Balakrishnan, Eur. Phys. J. B {\bf 15}, 679 (2000). 
 \bibitem{Stiller} J. C. Stiller and G. Radons, Phys. Rev. E {\bf 58},
	 1789(1998). 
 \bibitem{Daido} H. Daido, Phys. Rev. Lett. {\bf 68}, 1073 (1992).
 \bibitem{Theumann} W. K. Theumann, Physica A {\bf 328}, 1(2003). 
 \bibitem{Diederich} S. Diederich and M. Opper, Phys. Rev. A {\bf 39}, R4333(1989).
 \bibitem{Wigner} E. P. Wigner, Ann. Math. {\bf 62}, 548 (1955).
\end{thebibliography}
\end{document}